\renewcommand{\thefootnote}{\fnsymbol{footnote}}

\documentstyle[11pt]{article}

\begin{document}

\topmargin 0pt
\oddsidemargin 5mm

\begin{titlepage}
\setcounter{page}{0}
\rightline{Preprint YERPHY-1428(15)-94}

\vspace{15mm}
\begin{center}
{\Large {\bf ON THE POSSIBILITIES TO STUDY THE EXOTIC STATE $1^{-+}$ IN THE
EXPERIMENTS ON PHOTOPRODUCTION OF THE $\pi\eta,\pi\eta'$ SYSTEMS}}
\vspace{1cm}

{\large I.G. AZNAURYAN} \\
\vspace{1cm}
{\em Yerevan Physics Institute,  Republic of Armenia}\\
{(Alikhanian Brothers St. 2,  Yerevan 375036,  Armenia)}
\end{center}

\vspace{5mm}
\centerline{{\bf{Abstract}}}

We show that in photoproduction experiments there  are propitious conditions
to study the exotic state $1^{-+}$
in  the  $\pi\eta$  and $\pi\eta'$ channels. We show that
for unpolarized photon beam the
contributions with natural  and unnatural parity exchanges do not interfere
with each other, the fact which gives additional arguments in
the estimation of the correctness of phase shift analysis. In the
reaction of photoproduction of $\pi^+\eta$ system in the mass
range of $a_2(1320)$ more strong limit on the upper bound of
the product $\Gamma(1^{-+}\rightarrow \pi\gamma)\times{\rm
Br}(1^{-+}\rightarrow\pi\eta)$ can be obtained than in
experiments on Primakoff production on nuclei. In the
photoproduction of  $\pi^0\eta$ and $\pi^0\eta'$
systems there are propitious conditions to study the state
$1^{-+}$ in the coherent photoproduction on  nuclei, the fact
which will allow to enlarge significantly the statistics and makes
the phase shift analysis more simple.
\vspace{15mm}

November 1994

\end{titlepage}

\renewcommand{\thefootnote}{\arabic{footnote}}
\setcounter{footnote}{0}

\setcounter{page}{1}

\section {Introduction}
\indent

It is known that many years much effort has been devoted to
the experimental search for exotic states, experiments being
mostly focused on meson spectroscopy. Number of mesons are
observed: $f_0(975), a_0(980),\eta/i
(1460),\Gamma/f_0(1590)\ldots$ which are candidates for exotics,
however by quantum numbers they can be ordinary
$q\bar{q}$-states, and there are doubts about unambiguous
identification of these  mesons as exotic states. Apart from
non-$q\bar{q}$ quantum numbers there is no simple unambiguous way
of distinguishing $q\bar{q}$ from non-$q\bar{q}$ objects. By this
reason it is of great interest to search for the state with
quantum numbers $J^{PC}=1^{-+}$ which does not fit into $q\bar{q}$
classification. The  most convenient channels to search for this
state are systems $\pi\eta,\pi\eta',\eta\eta$... which by selection rules
have limited number of states $0^{++},1^{-+},2^{++}\cdots,$ and
observation of the resonance in the P-wave for these systems
would provide unambiguous evidence for exotic. It should be
mentioned, however, that from  the point of   view of some
existing models (flux-tube model \cite{IHP} and nonrelativistic
potential model \cite{IYO}) these channels are not favourable for
the search of exotic state $1^{-+}$. But, strictly speaking, we are
unable currently to  obtain strict results in this field, and for
understanding of QCD in  non-perturbative sector complex
investigations are necessary with search for exotic states in
all channels.

In the $\pi\eta$ and $\pi \eta'$ systems at present there are
some indications on the possible existence of the exotic state
$1^{-+}$ \cite{DA}-\cite{YG}. The data are obtained on the pion beams
in the following experiments: by the GAMS collaboration at
CERN-SPS  in the charge-exchange reaction $\pi^- p
\rightarrow\pi^0\eta n$ at $100 {\rm GeV} \cite{DA}$, in the
diffractive reactions $\pi^- p \rightarrow\pi^-\eta p$ and $\pi^- p
\rightarrow\pi^-\eta'p$ at KEK at 6.3 ${\rm GeV} \cite{HA}$
and on the VES detector at IHEP at 37 ${\rm GeV} \cite{YG}$.
However, the data are not sufficient to  make unambiguous phase
shift analyses in these experiments and new experiments
including new reactions are neccessary.

In this paper we will show that in photoproduction experiments
there are propitious conditions to study the $1^{-+}$state in the
$\pi\eta$ and $\pi\eta'$ channels. In section 2 we will
show that for unpolarized photon beam the formulas for angular
distributions of  pions (or $\eta$-mesons)
in the Gottfried-Jackson frame  of $\pi\eta$ system are the same
as for  pion beam \cite{HA,GC}. As in the case of pion beam the
mechanisms with natural parity exchange (NPE) and unnatural parity
exchange (UPE) in the t-channel
do not interfere with each other and contribute to different amplitudes
with different angular dependences.
This fact is usefull in
order to find the mechanism of production of the resonance
and can give additional arguments in the estimation of the
correctness of phase  shift analysis. In
section 3 we will show that in reaction of photoproduction of
$\pi^+\eta$ system in the mass range of $a_2(1320)$-resonance more
strong limit on the upper bound of the product
$\Gamma(1^{-+}\rightarrow\pi\gamma)\times{\rm Br}(1^{-+}\rightarrow
\pi\eta)$ can be obtained than in the experiment on Primakoff
production on nuclei \cite{MZ}. In section  4 we will show that
photoproduction of $\pi^0 \eta $ and $\pi^0 \eta'$ systems at
small momentum transfers is dominated by spin-nonflip
zero-isospin NPE. This means that  there are propitious
conditions to study the state $1^{-+}$ in the coherent
photoproduction on nuclei, the  fact which will allow to enlarge
significantly the statistics and makes the phase shift analysis
more simple, as in this case only contributions  with NPE
exchange survive.

\section{Cross section expansion over helicity
states of the $\pi\eta$ system}
\indent

In this section we will carry out parallel consideration of the
reactions
\begin{equation}
\label{AB}
\pi N \rightarrow(\pi\eta)N,
\end{equation}
\begin{equation}
\label{AC}
\gamma N \rightarrow(\pi\eta)N
\end{equation}
and will show that for pion beam and for unpolarized photon beam
formulas for angular distributions in the Gottfried-Jackson
frame of the $\pi\eta$  system coincide.

Let us define helicity amplitudes for these reactions as
\begin{equation}
\label{AD}
M^\pi(\lambda,\lambda_f,\lambda_i)\quad and \quad M^\gamma(\lambda,\lambda_f,
\lambda_\gamma,\lambda_i),
\end{equation}
where $\lambda$ is helicity of $\pi\eta$
system, $\lambda_i$ and $\lambda_f$ are helicities of initial and final
nucleons, $\lambda_\gamma$ is  photon  helicity. For a definite parity exchange
(natural or unnatural) in the t-channel there is a parity relation at each
vertex \cite{GCT} which gives in the verteces $\pi \rightarrow\pi\eta$ and
$\gamma \rightarrow\pi\eta$:
\begin{equation}
\label{AE}
M^\pi(-\lambda,\lambda_f,\lambda_i)=P \eta_\pi\eta_{\pi\eta}
(-1)^{L+\lambda} \, M^\pi(\lambda,\lambda_f,\lambda_i),
\end{equation}
\begin{equation}
\label{BC}
M^\gamma(-\lambda,\lambda_f,-\lambda_\gamma,\lambda_i)=P \eta_\gamma\eta_
{\pi\eta}(-1)^{L+\lambda-S_\gamma-\lambda_\gamma}\,
M^\gamma(\lambda,\lambda_f,\lambda_\gamma,\lambda_i).
\end{equation}
Here $P=+1$ for NPE and $P=-1$ for
UPE, $\eta$ denotes internal parity of the states, $S_\gamma$ and $L$
are spins of photon and $\pi\eta$ system.
For the $\pi\eta$ system $\eta_{\pi\eta}=(-1)^L$,
so we have
\begin{equation}
\label{CD}
M^\pi(-\lambda,\lambda_f,\lambda_i)=-P(-1)^\lambda\, M^\pi(\lambda,
\lambda_f,\lambda_i),
\end{equation}
\begin{equation}
\label{DE}
M^\gamma(-\lambda,\lambda_f,-\lambda_\gamma,\lambda_i)=-P(-1)^\lambda
\, M^\gamma(\lambda,\lambda_f,\lambda_\gamma,\lambda_i).
\end{equation}

Now we can define amplitudes with definite parity exchange in the t-channel:
\begin{equation}
\label{EA}
M^\pi_{\lambda\pm}\equiv M^\pi_\lambda\mp(-1)^\lambda M^\pi_{-\lambda},
\end{equation}
\begin{equation}
\label{EC}
M^\gamma_{\lambda\pm}\equiv M^\pi_{\lambda,\lambda_\gamma}\mp(-1)^\lambda
M_{-\lambda,-\lambda_\gamma}.
\end{equation}

If we consider the reactions with unpolarized photons and nucleons and,
therefore, sum up over their helicity states,
the relations of  type of eq.(\ref{EA}) can be written
for  each indeces of the density matrices of reactions (\ref{AB}) and
(\ref{AC}):
\begin{eqnarray}
\label{GG}
&&\rho_{\lambda,\lambda^\prime}=\frac{1}{2}(\rho_{\lambda+,\lambda^\prime}
+\rho_{\lambda-,\lambda^\prime}),\nonumber \\
&&\rho_{-\lambda,\lambda^\prime}=\frac{-(-1)^\lambda}{2}(\rho_{\lambda+,\lambda^
\prime}-\rho_{\lambda-,\lambda^\prime}),\nonumber \\
&&\rho_{\lambda,\lambda^\prime}=\frac{1}{2}(\rho_{\lambda,\lambda^\prime+}
+\rho_{\lambda,\lambda^\prime-}),\\
&&\rho_{\lambda,-\lambda^\prime}=\frac{-(-1)^{\lambda^\prime}}{2}(\rho_
{\lambda,\lambda^\prime+}-\rho_{\lambda,\lambda^\prime-}).\nonumber
\end{eqnarray}

These density matrices define the angular distributions of $\pi$ and $\eta$
in the  Gottfried-Jackson frame of the $\pi\eta$ system  as follows:
\begin{equation}
\label{GA}
\frac{d\sigma}{d\Omega}\sim W(\Theta,\varphi)=
\sum_{L,L^\prime}\sum_{\lambda,\lambda^
\prime}{\left(\frac{2L+1}{4\pi}\,
\frac{2L^\prime+1}{4\pi}\right)}^{1/2}\,
D_{\lambda0}^L(\Theta,\varphi)\quad [D^{L^\prime}_{\lambda^\prime0}
(\Theta,\varphi)]^* \rho_{\lambda\lambda^\prime},
\end{equation}
where $\Theta$ and $\varphi$ are polar and azimutal angles of $\pi$ (or $\eta$)
in this frame. Now using relations (10), the relations
\begin{equation}
\label{GB}
D_{-|\lambda| 0}^L=(-1)^\lambda\,(D_{|\lambda| 0}^L)^*,\quad
D_{\lambda0}^L(\Theta,\varphi)=e^{i\lambda\varphi} d_{\lambda 0}^L (\Theta)
\end{equation}
and the symmetry of $W(\Theta,\varphi)$
according to  the replacement $\varphi\rightarrow-\varphi$ we can write
$\displaystyle
\frac{d\sigma}
{d\Omega}$ in the form \cite{HA,GC}:
\begin{equation}
\label{FF}
\frac{d\sigma}{d\Omega}=|A_0+A_-|^2+|A_+|^2,
\end{equation}
where
\begin{eqnarray}
\label{FC}
A_0&=&\sum_{L=0}^{L_{\rm max}}\left(2L+1\right)^{1/2}L_0 D_{00}^L(\Theta,
\varphi),\nonumber\\
A_-&=&\sum_{L=0}^{L_{\rm max}}\sum_{\lambda=1}^{L}\left(2L+1\right)^{1/2}
\sqrt{2}L_{\lambda -} {\rm Re}(D_{\lambda 0}^L(\Theta,\varphi)),\\
A_+&=&\sum_{L=0}^{L_{\rm max}}\sum_{\lambda=1}^{L}\left(2L+1\right)^{1/2}
\sqrt{2}L_{\lambda +} {\rm Im}(D_{\lambda 0}^L(\Theta,\varphi)).\nonumber
\end{eqnarray}

Here we have  used notations of ref.[4], $L_0$ and $L_{\lambda-}$
are amplitudes which correspond to production of
$\pi \eta$ system with spin $L$ via UPE,
$L_{\lambda+}$ via NPE. In each bilinear product of amplitudes
a summation  over photon and nucleon helicities is implicit.

 From the derivation of the formulas  (\ref{FF}),(\ref{FC}) it is seen that
the similar relations can  be written in the case of
electroproduction for cross sections $\sigma_t$ and $\sigma_s$
corresponding to transverse and longitudinal virtual photons. For
$\sigma_t$ we will have (as in  the case of  unpolarized photon
beam) that $L_0$ and $L_{\lambda-}$ correspond to  UPE and
$L_{\lambda+}$ to NPE. However, for longitudinal photons the sign
in right hand  side of eq.(\ref{DE}) will be positive. By this
reason for $\sigma_s$ amplitudes $L_0$ and $L_{\lambda-}$
will correspond to NPE and $L_{\lambda+}$ to UPE.

Let us emphasize the following feature of the relations
(\ref{FF}),(\ref{FC}): in the considered reactions the mechanisms
with NPE and UPE lead to quite different angular distributions in
the Gottfried-Jackson frame of $\pi\eta$ system and contribute to
different amplitudes. This fact can be used as additional argument in the
estimation of correctness of the  results of the phase shift
analyses. Let us demonstrate this on the phase shift analyses made
in refs.\cite{DA}-\cite{GC}.

According to the phase shift analysis of the reaction $\pi^-
p \rightarrow \pi^0\eta n$ made in ref. \cite{DA} the exotic
state $1^{-+}$ appears in the amplitude $P_0$, i.e. its
production mechanism corresponds to UPE. At the same time at the
energies used in this experement (100 GeV) the main contributions
should be $\rho$-exchange with NPE and $\rho P$-cuts with
predominantly NPE. The consideration as the production mechanism
of the $b_1(1230)$-exchange \cite{SFT} with UPE has low probability
at these energies due to low intercept of this trajectory. By
these reasons the results of the phase shift analysis of Alde et al
\cite{DA} are doubtfull.

The situation is different in the ref.\cite{HA,YG}. Here the
diffractive reactions $\pi^- p\rightarrow\pi^-\eta(\eta') p$ with
predominantly NPE are investigated. According to the phase shift
analyses made in \cite{HA,YG} the possible exotic state
$1^{-+}$ appears in the $P_+$ amplitude, and so, from the point of
view of correctness of  the production mechanism these
analyses are out of  doubt.

In order to fix a normalization in eqs.(\ref{FF}),(\ref{FC}) let us give
formulas
for dioganal terms corresponding to  the production of $\pi\eta$ system
through the resonance state with spin $L$:
\begin{equation}
\label{FD}
d\sigma= d\sigma_L\frac{\frac{1}{\pi}m_L\Gamma_L dM^2}{(M^2-m_L^2)^2+
(m_L\Gamma_L)^2}Br(\eta\pi)W_L ({\rm cos}\Theta)d {\rm cos}\Theta,
\end{equation}
where $d\sigma_L$ is the cross  section of production of the resonance,
M is invariant  mass of $\pi\eta$ system:
\begin{equation}
\label{FN}
M^2=(P_\pi+P_\eta)^2,
\end{equation}
$m_L$ and $\Gamma_L$ are the resonance mass and total width, $W_L({\rm
cos}\Theta)$
is integrated over azimutal angle angular distribution of $\pi$(or $\eta$)
in the Gottfried-Jackson frame. For unpolarized photon beam and for pion beam
this angular distribution has the form:
\begin{eqnarray}
\label{FM}
W_0({\rm cos}\Theta)&=&\frac{1}{2}\, ,\nonumber\\
W_1({\rm cos}\Theta)&=&\frac{3}{2}\left[\rho_{00}^U{\rm cos}^2\Theta+
(\rho_{11}^N+\rho_{11}^U){\rm sin}^2\Theta\right],\\
W_2({\rm cos}\Theta)&=&\frac{5}{8}\left[\rho_{00}^U(3{\rm cos}^2\Theta-
1)^2+12(\rho_{11}^N+\rho_{11}^U){\rm sin}^2\Theta{\rm cos}^2\Theta+
3(\rho_{22}^N+\rho_{22}^U){\rm sin}^4\Theta\right],\nonumber
\end{eqnarray}
were ${\rho_{LL}^{U,N}}$-are density matrix elements corresponding
to UPE and NPE. They obey the conditions:
\begin{eqnarray}
\label{FX}
{\rm for}\,\, L=1:\,\,\rho_{00}^U+2(\rho_{11}^N+\rho_{11}^U)=1,\nonumber\\
{\rm for}\,\, L=2:\,\,\rho_{00}^U+2(\rho_{11}^N+\rho_{11}^U+\rho_{22}^N+
\rho_{22}^U)=1.
\end{eqnarray}

\section{Photoproduction of the $\pi^+ \eta $ system}
\indent

In this section we will show that
in the reaction of photoproduction of the $\pi^+\eta$ system in
the mass range of the resonance $a_2(1320)$ it is possible
to obtain more strong limit on the upper bound of the
product $\Gamma(1^{-+}\rightarrow\pi\gamma)
{\rm Br(1^{-+}\rightarrow\pi\eta)}$ than in the experiment on
Primakoff production \cite{MZ}. Experimental data
\cite{YE} indicate on the  large cross section of
photoproduction of $a_2(1320)$ in the  energy range 3-6 GeV,
and the resonance $a_2(1320)$ has large branching ratio into
$\pi\eta$ channel $=14.5\pm 1.2\%$. From energy depedence and
t-distributions of the data it follows that photoproduction of
$a_2^+(1320)$ is dominated by $\pi$-exchange and, hence, is
discribed  by the diagram of  fig.1a. In the case of possible
production of the positive resonance state $1^{-+}$ in the
reaction (\ref{AC}) it will be also dominanted by $\pi$-exchange
(the diagram of fig.1a). So, the photoproduction of the resonances
$a_2(1320)$ and $1^{-+}$ is determined by the same verteces as in
the Coulomb production of these resonances in the experiments
with pion beams (the diagram of fig.1b). These verteces can be
written in the form:
\begin{equation}
\label{II}
V(1^{-+}\rightarrow\pi\gamma)=g\varepsilon_{\mu\nu\sigma\lambda}
\varepsilon^\mu_{1^{-+}}\varepsilon_\gamma^\nu p_{1^{-+}}^\sigma
p_\pi^\lambda,
\end{equation}
\begin{equation}
\label{II1}
V(a_2\rightarrow\pi\gamma)=G \varepsilon_{\mu\nu\sigma\lambda}
\varepsilon^{\mu\alpha}_{a_2}\varepsilon_\gamma^\nu p_{a_2}^\sigma
p_\pi^\lambda(p_\pi-p_\gamma)_\alpha,
\end{equation}
where $\varepsilon$ and $p$ denote wave functions and four-momenta of
particles. Using these verteces it is easy
to show that at large energies and small momentum
transfers the  ratios of cross sections of the $a_2$ and $1^{-+}$
production via the diagrams of figs.1a and 1b coincide, being
equal to
\begin{equation}
\label{W1}
\frac{d\sigma(1^{-+})}{d\sigma(a_2)}=\frac{3\Gamma(1^{-+}\rightarrow
\pi\gamma)}{5\Gamma(a_2\rightarrow\pi\gamma)}.
\end{equation}

For convenience we have supposed here that
$M^2=m^2_{a_2}=[m(1^{-+})]^2$. The density matrix elements for
the diagrams of figs.1a and 1b also coincide with each other and
are equal to
$\rho_{00}=\rho_{22}=\rho_{11}^ N=0,\quad \rho_{11}^U=\frac{1}{2}$.
Therefore, using relations (\ref{FD}) and (\ref{FM}) we have for the
cross sections of production of the $\pi\eta$ system in the considered
reactions
the following expression:
\begin{equation}
\label{DD1}
d\sigma=d\sigma(a_2){\rm Br}(a_2\rightarrow\eta\pi)W(cos\theta),
\end{equation}
where
\begin{equation}
\label{DD2}
W(cos\theta)=\frac{15}{4}sin^2\theta\left(cos\theta\pm\sqrt{\Delta}
\right)^2,
\end {equation}
\begin{equation}
\label{DG}
\Delta=\frac{3\Gamma(1^{-+}\rightarrow\pi\gamma){\rm
Br}(1^{-+}\rightarrow\pi\eta)}{25\Gamma(a_2\rightarrow\pi\gamma){\rm
Br}(a_2\rightarrow\pi\eta)}.
\end{equation}

Let us note that sings in bracket in eq.(\ref{DD2}) are opposite to each other
for the diagrams of figs.1a and 1b due to the term
$(p_\pi-p_\gamma)_\alpha$ in (\ref{II1}).

So, the   possible existence of the  resonance $1^{-+}$ in the mass range
of $a_2(1320)$ will give the forward-backward asymmetry in the
Gottfried-Jackson frame of the $\pi\eta$ system which should be the
same in  magnitude (but opposite in sign) in the photoproduction
experiments and in  the Primakoff production experiments. In the
photoproduction experiments there are
conditions to study this asymmetry with better statistics than
in the experiments on Primakoff production. For example,
for the CLAS of CEBAF  the detected rate of $a_2^+$ in
the $\pi^+\eta$ channel will be $700/h$ assuming a $10{\rm
cm}\,LH_2$ target and a photon beam with intensity of $2\times10^7/s$.
\footnote{I am thankfull to S.S.Stepanyan  who has made  this estimation}
In this estimation the experimental data on $a_2^+$-photoproduction
obtained in \cite{YE} are used. So, the statistics in
photoproduction experiment will be much larger than in the Primakoff
production experiment \cite{MZ}, where the number of events with
$a_2$ production in the $\pi\eta$ chanell is 144.

\section{Photoproduction of the $\pi^0\eta\,(\pi^0\eta^\prime)$
systems}
\indent

Photoproduction of the exotic state $1^{-+}$ in the
$\pi^0\eta,\pi^0\eta'$
channels can go only through the C-odd exchanges $\omega$ and
$\rho$ (see diagrams of figs.2a,b), $\omega$ exchange being
dominating at small momentum transfers by the following reasons:

$(a)$ the coupling constant $g_{\omega pp}$ corresponding to the
spin-nonflip forward scattering (which is dominating at small momentum
transfers) is significantly larger than
similar $g_{\rho pp}$ constant: $g_{\omega pp\simeq}4g_{\rho pp}$ \cite{OD};

$(b)$ the coupling constant $g_{\rho\gamma}$ which according to  VDM
determines the diagram with $\omega$-exchange is larger than the
coupling constant  $g_{\omega\gamma}$ in the diagram with
$\rho$-exchange: $g_{\rho\gamma}\simeq 3g_{\omega\gamma}$.

So, photoproduction of the $\pi^0\eta$ and  $\pi^0\eta'$ systems is
dominated at small momentum transfers by spin-nonflip NPE with
zero isospin and its differential cross section is determined
mainly by the amplitude $A_+$ in (\ref{FF}). This means that
there are propitious conditions to study the state $1^{-+}$ in
the coherent photoproduction of the $\pi^0\eta$, $\pi^0\eta^\prime$
systems on the nuclei which has the following advantages:

(a) coherent  photoproduction on nuclei allow  to enlarge
significantly the statistics;

(b) it supresses strongly the
contributions of amplitudes $A_0$ and $A_-$ with UPE, as in this
case the cross section will be determined by the following
combinations of nucleon amplitudes:
\begin{equation}
\label{MM}
M_{\frac{1}{2}\frac{1}{2}}+M_{-\frac{1}{2}-\frac{1}{2}},
\end{equation}
which according to parity relation \cite{GCT} at nucleon vertex
(like to the relations (\ref{AE}),(\ref{BC}))
is equal to zero for UPE. So, in coherent
photoproduction only contribution of $A_+$ survives, the fact
which makes the phase shift analyses much more easier.

The background conditions in  photoproduction of  $1^{-+}$ state
in $\pi^0\eta(\pi^0\eta')\rightarrow\,4\gamma$ channels are also
propitious. The search for $1^{-+}$ in these channels leads to
the suppression of the background processes in forward direction,
because in this case the positive C-parity of the state  is fixed
and by this reason possible background processes via P-and $\pi$
exchanges, which are large in forward direction, are rejected.

\section{Acknowledgments}
\indent

I express my gratitude to N.Ya.Ivanov and S.S.Stepanyan for
useful discussions.

The research described in this publication was made possible in part
by Grant \# RYE 000 from the International Science foundation and by
INTAS  Grant \# 93-283.

\newpage

\begin{center}
Figure captions
\end{center}

Fig.1 The diagrams  corresponding to the production  of the
$\pi\eta, \pi\eta'$ systems through the resonance states $a_2$
and $1^{-+}$ in  the  reactions  $\gamma p\rightarrow\pi^+\eta(\eta') n$
(via  one-pion exchange) and $\pi^-p\rightarrow\pi^-\eta(\eta')
p$ (via Primakoff effect).

Fig.2 The diagrams  corresponding to the production  of the
$\pi\eta, \pi\eta'$ systems in  the  reaction  $\gamma N\rightarrow
\pi^0\eta(\eta') N$.

\newpage

   \input FEYNMAN
   \textheight 110mm \textwidth 159mm
   \begin{picture}(20000,10000)(0,0)
   \drawline\photon[\E\REG](-9000,9000)[10]                          
   \put(\pmidx,10000){$\gamma$}
   \put(\photonbackx,\photonbacky){\circle*{300}}            
   \drawline\scalar[\E\REG](\photonbackx,\photonbacky)[3]       
   \put(\particlemidx,9800){$a_2,1^{-+}$}
   \put(\scalarbackx,\scalarbacky){\circle*{300}}               
   \drawline\scalar[\SE\REG](\scalarbackx,\scalarbacky)[3]      
   \put(\particlebackx,14000){$\pi^+$}
   \drawline\scalar[\NE\REG](\scalarfrontx,\scalarfronty)[3]    
   \put(\particlebackx,4000){$\eta (\eta')$}
   \drawline\scalar[\S\REG](\photonbackx,\photonbacky)[4]
   \put(\scalarbackx,-2000){(a)}
   \put(2000,\pmidy){$\pi^{+}$}
   \drawline\fermion[\E\REG](\scalarbackx,\scalarbacky)[10500]
   \put(\particlemidx,\particlemidy){\vector(1,0){2}}
   \put(\particlemidx,-500){n}
   \drawline\fermion[\W\REG](\scalarbackx,\scalarbacky)[10000]
   \put(\particlemidx,\particlemidy){\vector(1,0){2}}
   \put(\particlemidx,-500){p}
   \put(\particlemidx,-500){p}
   \drawline\scalar[\E\REG](19000,8810)[4]                             
   \put(\particlemidx,9800){$\pi^{-}$}
   \put(\scalarbackx,\scalarbacky){\circle*{300}}
   \drawline\photon[\S\REG](\scalarbackx,\scalarbacky)[8]
   \put(28000,\pmidy){$\gamma$}
   \put(\scalarbackx,-2000){(b)}
   \put(\scalarbackx,-2000){(b)}
   \drawline\scalar[\E\REG](\photonfrontx,\photonfronty)[3]
   \put(\particlemidx,9600){$a_2,1^{-+}$}
   \put(\scalarbackx,\scalarbacky){\circle*{300}}
   \drawline\scalar[\SE\REG](\scalarbackx,\scalarbacky)[3]
   \put(\particlebackx,13800){$\pi^-$}
   \drawline\scalar[\NE\REG](\scalarfrontx,\scalarfronty)[3]
   \put(\particlebackx,3800){$\eta (\eta')$}
   \drawline\fermion[\E\REG](\photonbackx,\photonbacky)[10450]
   \put(\particlemidx,\particlemidy){\vector(1,0){2}}
   \put(\particlemidx,-500){p}
   \drawline\fermion[\W\REG](\photonbackx,\photonbacky)[8200]
   \put(\particlemidx,\particlemidy){\vector(1,0){2}}
   \put(\particlemidx,-500){p}
   \end{picture}
   \vskip 0.5in
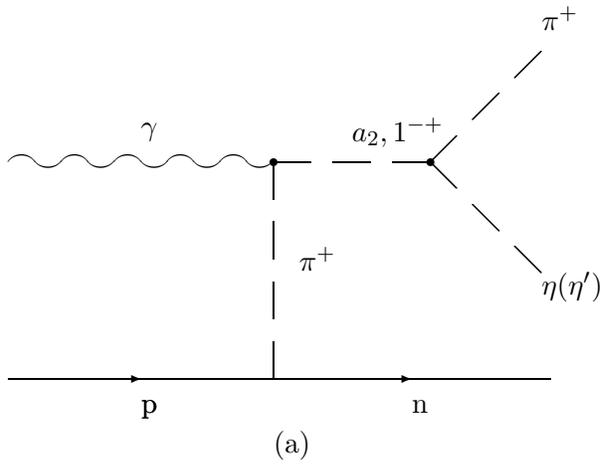
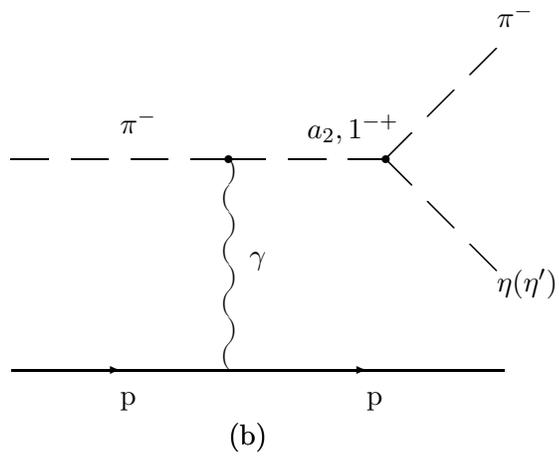
   \centerline{Fig.1}

   \textheight 110mm \textwidth 159mm
   \begin{picture}(10000,10000)(0,0)

   \drawline\photon[\E\REG](-9000,-5000)[9]                          
   \put(\pmidx,-4000){$\gamma$}
   \put(\photonbackx,\photonbacky){\circle*{300}}
   \drawline\scalar[\E\REG](\photonbackx,\photonbacky)[3]
   \put(\pmidx,-4000){$\rho^0$}
   \put(\scalarbackx,\scalarbacky){\circle*{300}}
   \put(\photonbackx,-16500){(a)}
   \drawline\scalar[\SE\REG](\scalarbackx,\scalarbacky)[3]
   \put(\particlebackx,-100){$\pi^0$} 
   \drawline\scalar[\NE\REG](\scalarfrontx,\scalarfronty)[3]
   \put(\particlebackx,-10000){$\eta (\eta')$}
   \drawline\scalar[\S\REG](\scalarfrontx,\scalarfronty)[4]
   \put(7000,\particlemidy){$\omega$}
   \drawline\fermion[\E\REG](\scalarbackx,\scalarbacky)[6500]
   \put(\particlemidx,\particlemidy){\vector(1,0){2}}
   \put(\particlemidx,-14000){p}
   \drawline\fermion[\W\REG](\scalarbackx,\scalarbacky)[7500]
   \put(\particlemidx,\particlemidy){\vector(1,0){2}}
   \put(\particlemidx,-14000){p}
   \drawline\photon[\E\REG](19000,-5000)[9]                           
   \put(\pmidx,-4000){$\gamma$}
   \put(\photonbackx,\photonbacky){\circle*{300}}
   \drawline\scalar[\E\REG](\photonbackx,\photonbacky)[3]
   \put(\particlemidx,-3900){$\omega$}
   \put(\scalarbackx,\scalarbacky){\circle*{300}}
   \drawline\scalar[\SE\REG](\scalarbackx,\scalarbacky)[3]
   \put(\particlebackx,-100){$\pi^0$} 
   \drawline\scalar[\NE\REG](\scalarfrontx,\scalarfronty)[3]
   \put(\particlebackx,-10000){$\eta (\eta')$}
   \drawline\scalar[\S\REG](\scalarfrontx,\scalarfronty)[4]
   \put(35000,\particlemidy){$\rho$}
   \put(\photonbackx,-16500){(b)}
   \drawline\fermion[\E\REG](\scalarbackx,\scalarbacky)[6500]
   \put(\particlemidx,\particlemidy){\vector(1,0){2}}
   \put(\particlemidx,-14000){p}
   \drawline\fermion[\W\REG](\scalarbackx,\scalarbacky)[7500]
   \put(\particlemidx,\particlemidy){\vector(1,0){2}}
   \put(\particlemidx,-14000){p}
   \end{picture}
   \vskip 2.7in
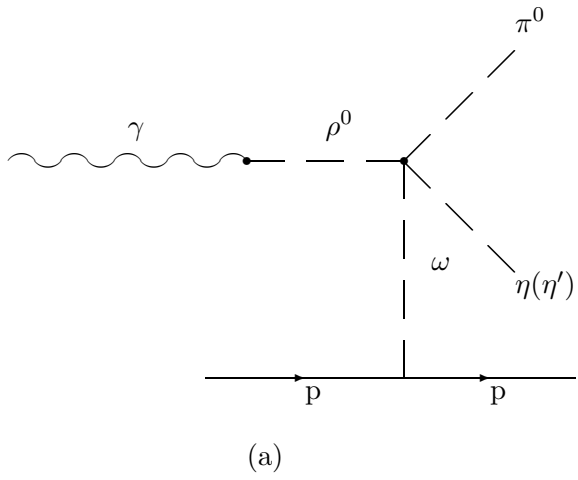
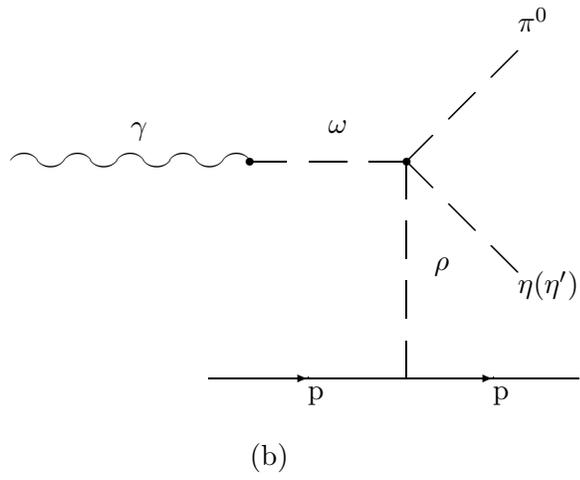
   \centerline{Fig.2}


\begin{thebibliography}{99}

\bibitem{IHP} N.Isgur, R.Kokoski, J.Paton, Phys.Rev.Lett. {\bf 54}
(1985) 869
\bibitem{IYO} F.Iddir, A.Le Yaouanc, L.Oliver et al., Phys.Lett. {\bf B205}
(1988) 564
\bibitem{DA} D.Alde et al., Phys.Lett. {\bf B205} (1988) 397
\bibitem{HA} H.Aoyagi et al., Phys.Lett. {\bf B314} (1993) 246
\bibitem{YG} G.M.Beladidze et al., Phys.Lett. {\bf B313} (1993) 276
\bibitem{GC}  G.Costa  et al., Nucl.Phys. {\bf B175} (1980) 402
\bibitem{MZ}  M.Zielinsky et al., Proc. XXIII Intern. Conf. on High Energy
Physics, Vol.1 ed.S.Lohen (World Scientific, Singapore, 1986) p.736;
S.Cihangir et al., Phys.Lett. {\bf B117} (1982) 119
\bibitem{GCT}G.Cohen-Tannoudji et al., Nuovo Cim. {\bf 55A} (1968) 412
\bibitem{SFT}S.F.Tuan, T.Ferbel, R.H.Dalitz, Phys.Lett. {\bf B213} (1988) 537
\bibitem{YE} Y.Eisenberg et al. Phys.Rev.Lett. {\bf 23} (1969) 1322;
Phys.Rev. {\bf D5} (1972) 15
\bibitem{OD} O.Dumbrajs et al., Nucl.Phys. {\bf B216} (1983) 277

\end{thebibliography}
\end{document}